\documentclass[12pt,preprint]{aastex}

\slugcomment{To be submitted to AJ, Draft \today (MWM)}

\shorttitle{2MASS J1707-0558AB}
\shortauthors{McElwain \& Burgasser}

\begin{document}

\title{Resolved Spectroscopy of M Dwarf/L Dwarf Binaries.
II. 2MASS J17072343-0558249AB}

\author{Michael W.\ McElwain\altaffilmark{1}}
\affil{Department of Physics \& Astronomy, University of
California at Los Angeles, Los Angeles, CA, 90095-1562, USA;
mcelwain@astro.ucla.edu}
\and
\author{Adam J.\ Burgasser\altaffilmark{1}}
\affil{Massachusetts Institute of Technology, 
Kavli Institute for Astrophysics and Space Research, 
77 Massachusetts Avenue, Building 37, Cambridge, 
MA 02139-4307, USA; ajb@mit.edu}

\altaffiltext{1}{Visiting Astronomer at the Infrared Telescope
Facility, which is operated by the University of Hawaii under
Cooperative Agreement NCC 5-538 with the National Aeronautics and
Space Administration, Office of Space Science, Planetary Astronomy
Program.}

\begin{abstract}

We present IRTF SpeX observations of the M/L binary system 2MASS
J17072343-0558249.  SpeX imaging resolves the system into a
1$\farcs$01$\pm$0.17 visual binary in which both components have red
near infrared colors.  Resolved low-resolution (R$\sim$150) 0.8--2.5
$\micron$ spectroscopy reveals strong H$_2$O, CO and FeH bands and
alkali lines in the spectra of both components, characteristic of
late-type M and L dwarfs.  A comparison to a sample of late-type field
dwarf spectra indicates spectral types M9 and L3.  Despite the small
proper motion of the system (0$\farcs$100$\pm$0$\farcs$009 yr$^{-1}$),
imaging observations over 2.5~yr provide strong evidence that the two
components share common proper motion.  Physical association is also
likely due to the small spatial volume occupied by the two components
(based on spectrophotometric distances estimates of 15$\pm$1 pc) as
compared to the relatively low spatial density of low mass field
stars.  The projected separation of the system is 15$\pm$3 AU, similar
to other late-type M and L binaries.  Assuming a system age of 0.5-5
Gyr, we estimate the masses of the binary components to be 0.072-0.083
and 0.064-0.077 M$_{\sun}$, with an orbital period of roughly 150-300 yr.
While this is nominally too long a baseline for astrometric mass
measurements, the proximity and relatively wide angular separation of
the 2MASS J1707-0558AB pair makes it an ideal system for studying the
M dwarf/L dwarf transition at a fixed age and metallicity.

\end{abstract}


\section{Introduction}

Wide field photographic imaging and proper motion surveys of the past
were largely incapable of producing an accurate census of nearby very
low mass (VLM; M $<$ 0.1~M$_{\sun}$) stars and brown dwarfs.
Late-type dwarfs emit weakly at visual bands, but are brighter at red
optical and near infrared wavelengths.  Progress in optical and
infrared detector technology have paved the way for the current
generation of wide field sky surveys, most notably the Deep Near
Infrared Sky Survey \citep{epc97}, the Sloan Digital Sky Survey
\citep{yor00} and the Two Micron All Sky Survey \citep[hereafter
2MASS]{skr06}.  Searches for faint red sources in these catalogs
(e.g., \citealt{del97,kir99,fan00}) have identified hundreds of very
low mass stars and brown dwarfs, greatly expanding our VLM census and
revealing the new spectral classes L \citep{kir99,mar99} and T
\citep{bur02,geb02}.

The nascent field of VLM stars has been confronted with a number of
fundamental questions regarding the properties of low luminosity
sources, including formation scenarios, thermal evolution, chemical
compositions and dynamics of cool stellar/substellar atmospheres, and
the initial mass function.  Multiple star systems are important
laboratories for understanding these physical properties.  Binary
systems are key in the determination of stellar masses, which can be
used to calibrate theoretical evolutionary and structure models.
Binary parameters probe the star formation process, and relative
comparisons can be made between coeval components.  Searches for cool
dwarf binaries through high resolution imaging have been conducted to
explore the nature of VLM stars and brown dwarfs (e.g.,
\citealt{mar99,koe99,rei01,clo02,clo03,bou03,bur03a,
giz03,sie05,law06,bur06c}).  The result of these efforts has been the
discovery of roughly 75 VLM binary systems with varying mass ratios
and projected separations (cf.\ Burgasser et al.\ 2006).

The properties of the current sample of VLM stars and brown dwarfs
suggests that both the multiplicity fraction and peak of the semimajor
axis distribution are directly related to the primary star mass.  G to
M binaries tend to peak at separations of 3-30 AU
\citep{fis92,hen93,rei97}, with binary fractions $f_{bin}$ ranging
from $\sim$65\% \citep{duq91} down to $\sim$30\%
\citep{fis92,rei04,del04}.  In contrast, nearly all VLM binaries have
separations $\rho~$$\lesssim$ 20 AU, with a binary fraction $f_{bin}$
$\approx$ 15\% (e.g., \citealt{clo03}; however, see also
\citealt{max05}).  The low frequency and preference of small
separations for VLM binaries is a challenge for star formation
theories, and a clear understanding of binary parameters as a function
of mass, age, and metallicity provides empirical clues for
understanding VLM star and brown dwarf formation processes.

Studies of VLM binaries also reveal the detailed properties of cool
star and brown dwarf atmospheres.  Dust formation and evolution
remains an outstanding problem across the spectral transition from M
dwarfs to L dwarfs (hereafter the M/L transition), where lower
temperatures and higher pressures enable some molecules to form solid
condensates and descend from the photosphere, changing the overall
morphology of the spectrum and the atmospheric pressure/temperature
profile \citep{lun89,tsu96,lod99,lod02,burr99,ack01,lod02b}.  In
particular, M dwarfs are characterized by their strong TiO and VO
absorption bands; but as temperatures descend into the L dwarf regime,
refractory elements such as Fe, Mg, Ti, V, Al, and Ca are removed from
the gas in the photosphere by the condensation and sedimentation
process.  Theoretical models examining these processes are compared to
the existing spectra of field dwarfs, which have a variety of
metallicities, masses, and ages.  M/L binaries, on the other hand, can
help clarify the existence of atmospheric condensation by isolating
parameters such as age and metallicity, assuming coevality.

A total of 14 binaries comprised of M and L dwarf components have been
discovered to date \citep[see Burgasser et al.\
2006]{giz00,giz03,bou03,clo03,fre03,mar03,sie03,sie05,cha04,cha05,bil05,bur06}.
This paper presents the discovery of a new M/L binary\footnote{Reid et
al.\ (2006) have concurrently resolved this system using the Hubble
Space Telescope NICMOS instrument.  Their photometric estimates of the
spectral types and distance of the 2MASS 1707-0558AB system agree with
our analysis.}, 2MASS J17072343-0558249AB (hereafter 2MASS 1707-0558),
identified via resolved near infrared imaging and spectroscopy using
the SpeX instrument \citep{ray03} mounted at the 3 m NASA Infrared
Telescope Facility (IRTF). In $\S$~2 we describe our observations and
data reduction, and present results.  In $\S$~3 we analyze these data,
determining individual spectral types for the resolved components and
the photometric properties of the system.  We argue for physical
association by common proper motion, similar spectrophotometric
distances, and comparing the volume occupied by the M/L pair to the
measured density of VLM dwarfs in the Solar Neighborhood.  We also
make preliminary estimates of the individual masses and orbital
characteristics of the system.  We discuss our results in $\S$~4,
placing the 2MASS 1707-0558 system in context with other VLM binaries.
This work is summarized in $\S$~5.

\section{Observations}

The unresolved source 2MASS 1707-0558 was discovered by \citet{giz02}
in a search for late-type dwarfs in the direction of the TW Hydrae
association using the 2MASS survey.  Optical spectroscopy of the
composite system infers a spectral type of M9 on the \citet{kir99}
late-M and L dwarf scale, and confirms this source as a normal field
dwarf.  H$\alpha$ emission, common for late-type M dwarfs
\citep{giz00,wes04}, was observed with an equivalent width of 0.4
{\AA}.  The signal-to-noise and resolution of the optical spectrum was
not sufficient to detect the 6708 {\AA} \ion{Li}{1} line, so this
source was suspected to be a nearby Hydrogen-burning star or brown
dwarf.

\subsection{Imaging}

2MASS 1707-0558 was first observed on 2003 March 23 (UT) as a spectral
comparison star for the 2MASS Wide Field T Dwarf Search program
\citep{bur03b}.  Observing conditions were good with 0$\farcs$5 seeing
at $J$-band, and the SpeX imager/guider was used to sample a
60$\arcsec$$\times$60$\arcsec$ field of view at 0$\farcs$12
pixel$^{-1}$.  While acquiring 2MASS 1707-0558 with the imager, we
resolved two sources at the given sky position.  We subsequently
obtained $J$, $H$ and $K$\footnote{$JHK$ filters for SpeX are based on
the Mauna Kea Observatories near-infrared (MKO-NIR) system
\citep{sim02,tok02}.} images of the visual pair for color comparison.
Integrations of 15 s were obtained in 4 dithered exposures on the
chip.  A second series of images was obtained on 2004 August 9, with
slightly hazy conditions and 0$\farcs$9 seeing at $J$. Images were
slightly out of focus during the second campaign.  Integrations of 20,
15, and 12 s with 4 dithers were obtained in the $J$, $H$, and
$K$-band, respectively.  We observed the single star USNO-A2.0
0825-10078125 \citep{mon98} concurrently with the 2MASS 1707-0558
observations to serve as a point spread function (PSF) calibrator.
Finally, a third epoch of images was obtained on 2005 August 10,
during poor conditions and 0$\farcs$9 seeing at $J$.  Integrations of
10 s with 4 dithers were obtained in both the $J$ and $K$-band.  In
2005, we observed the nearby star USNO-A2.0 0825-10079794 as a PSF
calibrator.  A log of our imaging observations is provided in Table 1.

All imaging data was reduced in a typical procedure for near infrared
imaging.  Flat fields were constructed with sky frames, which were
median combined, subtracted by a median combined dark frame of the
same integration time, and normalized.  An image mask was
constructed to distinguish deviant pixels that were excessively bright
in the dark frames and cold in the sky flats.  The target and PSF
images were pair-wise subtracted, divided by the normalized flat,
cleaned by linear interpolation over the bad pixels, and added
together by integer pixel shifts to match the peak flux position of
the brighter source.

Reduced $J$, $H$ and $K$-band images from the 2003 observations of
2MASS 1707-0558 are displayed in Figure 1.  The two sources are
clearly resolved in all three bands and separated by roughly
1$\arcsec$ on the sky.  We discuss the relative fluxes and astrometry
for the pair in $\S$~3.1.

\subsection{Spectroscopy}

Near-infrared spectra of both components were acquired on 2003 May 23
(UT) with SpeX in prism mode, using the 0$\farcs$5 slit.  This
configuration yields a single-order, low-resolution (R$\sim$150)
0.8--2.5 $\micron$ spectrum with a dispersion of 20--30 {\AA}
pixel$^{-1}$ onto the Aladdin 3 1024$\times$1024 InSb array.
Conditions during the observations were good with seeing of 0$\farcs$5
at $J$-band, and the target was observed at an airmass of 1.16.  In
order to obtain the spectrum of each component separately, we aligned
the slit perpendicular to the orientation of the two sources, guiding
on one component while observing the other through the slit.  Based on
the slit width, component separation and PSF full-widths at
half-maximum, we estimate that only 4-6\% of the light from the
guiding component contaminated the slit.  Exposures of 120/180s were
taken as dithered pairs for the brighter/fainter component.  The A0
star HD 171149 was observed immediately after the target exposures at
similar airmass (1.17), followed by internal flat-field and Ar arc
lamps for pixel response and wavelength calibration.

Data were reduced using the Spextool package version 3.2
\citep{cus04}.  The raw target data were processed by performing
linearity corrections, pair-wise subtraction, and division by a
normalized flat field.  The target spectra were then extracted using
the Spextool default settings for point sources, and wavelength
solutions were calculated using the Ar arc calibration frames.
Extracted spectra from the same source were scaled to match the
highest S/N spectrum of the set, and the scaled spectra were median
combined.  The resultant spectrum was removed of telluric features,
intrinsic A0V star lines, and instrumental response signatures
following the procedures of \citet{vac03}.  A Vega model spectrum was
employed to produce a true telluric spectrum from the A0 standard.
The model spectrum was shifted to match the radial velocity of the
standard spectrum, and then the modified model spectrum was reddened
and scaled to match the standard spectrum.  Line shape kernels were
constructed using unresolved arc lines from the wavelength calibration
observations.  These kernels were then used to broaden the model
spectrum line widths and smooth the model spectrum to match the
observed resolution.  The model spectrum was completed by adjusting
the H line strengths to reflect that of the observed A0 standard.  The
corrected telluric spectrum is constructed by dividing the model
spectrum by the A0 spectrum. The corrected telluric spectrum is then
multiplied by the target spectrum to produce the final flux-calibrated
spectrum.

The reduced spectra of the two components are plotted in Figure 2.
The spectrum of the bright component exhibits TiO absorption at 0.76,
0.82, and 0.84 $\micron$, VO absorption at 1.05 $\micron$, \ion{K}{1}
doublets at 1.17 and 1.25 $\micron$, the onset of FeH absorption at
0.98, 1.19, and 1.58 $\micron$, CO at 2.3 $\micron$, and strong H$_2$O
absorption at 1.4 and 1.9 $\micron$.  The spectrum of the faint component
has diminished TiO and VO absorption features, but still shows the
CrH, \ion{K}{1} doublets, FeH, CO and H$_2$O absorption features.  In
addition, the peak flux of the spectral energy distribution for the
faint component is also shifted redward to 1.28 $\micron$.  These
spectral features suggest that the bright component is a late M dwarf,
consistent with the optical spectral type of \citet{giz02}; while the
faint component shows features which are indicative of an early L
dwarf.  We discuss the classification of these sources in further
detail in $\S$~3.2.

\section{Analysis}

\subsection{PSF Fitting}

The PSF wings of the two sources at the position of 2MASS 1707-0558
are slightly blended in the 0$\farcs$5, 0$\farcs$9, and 0$\farcs$9
seeing of the 2003, 2004, and 2005 image data, respectively.  We
therefore determined the relative fluxes and astrometry of the pair
through a PSF fitting algorithm similar to that employed by
\citet{bur06}.  For the 2004 and 2005 images, we used a comparison
star as the PSF model.  For each filter, the components of the 2MASS
1707-0558 pair were first fit to a two-dimensional Gaussian to
determine the approximate flux center and a rough estimate of the peak
amplitude.  A synthetic image was created by shifting, scaling and
adding the corresponding PSF star images to match each of the
component positions and fluxes.  This model image was subtracted from
the original, and the standard deviation of the residual image was
used to measure the quality of the fit.  The model image was
iteratively modified, changing the position of the primary, the
position of the secondary, the peak flux of the primary and the peak
flux of the secondary, in that order, to improve the fit until a
minimum in the standard deviation of the residual image was achieved.
This routine was performed on every pair-wise subtracted image for
each filter in order to estimate the experimental uncertainty of the
fits.  For the 2003 data, which did not include a PSF comparison star,
we used the same algorithm with a fixed width Gaussian profile to
model the PSF.  The astrometric results --- measured separations
($\rho$) and position angle ($\phi$) -- are listed in Table
2. Relative magnitudes and aggregate astrometric measurements are
given in Table 3.  The uncertainties in all of these measurements
correspond to the 1$\sigma$ scatter in the values computed from all
image frames.

Relative magnitudes were measured using MKO-NIR filters, but composite
systemic photometry is on the 2MASS system.  In order to derive an
apparent magnitude for each component, we converted the MKO-NIR
relative magnitudes (${\Delta}M^{MKO}$) to the 2MASS system as
\begin{eqnarray}
{\Delta}M^{2MASS} & = & M_b^{2MASS}-M_a^{2MASS} \\
& = & M_b^{MKO}-M_a^{MKO}+(M_b^{2MASS}-M_b^{MKO})-(M_a^{2MASS}-M_a^{MKO}) \\
& = & {\Delta}M^{MKO}+\delta_{b}-\delta_{a},
\end{eqnarray}
where $\delta \equiv M^{2MASS}-M^{MKO}$ is the filter translation
factor.  These values were determined by calculating synthetic
magnitudes from the spectra of the 2MASS 1707-0558 components, using
MKO-NIR and 2MASS relative response curves ($R_M(\lambda)$ = filter
$M$ transmission function $\times$ optical response $\times$ telluric
absorption at an airmass of 1) from S.\ Leggett (2004, private
communication) and \citet{coh03}\footnote{See \citet{cut03},
$\S$~IV.4.a:
\url{http://www.ipac.caltech.edu/2mass/releases/allsky/doc/sec6\_4a.html\#rsr}.},
respectively:\footnote{Note that we do not include an additional
factor of $\lambda$/hc in Eqn.\ 5 for the MKO-NIR filters to
compensate for the photon counting properties of current detectors
(this factor is already included in the response curves of
\citet{coh03}).  This omission only leads to a $\sim$0.01 mag offset
in the derived magnitudes for late-type dwarf spectra \citep{ste04}.
We thank our referee for pointing out this detail.}
\begin{equation}
M = -2.5\log_{10}
\left(\frac{\int{f_{\lambda}(\lambda)R_X(\lambda)d\lambda}}{\int{f_{\lambda}^{Vega}(\lambda)R_M(\lambda)d\lambda}}
\right).
\end{equation}
Here, $f_{\lambda}^{Vega}$ is the observed spectrum of the A0V star
Vega from \citet{hay85}.  The filter translation factors were measured
to be $\delta_{J,b}$$-$$\delta_{J,a}$ = $-$0.026,
$\delta_{H,b}$$-$$\delta_{H,a}$ = 0.013, and
$\delta_{K,b}$$-$$\delta_{K,a}$ = 0.002.  Using these values,
individual component magnitudes were computed from the composite 2MASS
magnitudes for the 2004 and 2005 observations; mean values and their
errors are listed in Table 3.

\subsection{Spectral Classification}

Spectral classification of stars is traditionally performed by
comparison to a template of spectral standards, observed with the same
dispersion and signal-to-noise ratio \citep{mor43}.  Unfortunately, a
set of M and L dwarf spectral standards does not yet exist in the near
infrared, although methods of near infrared classification have been
considered by various groups
\citep{tok99,rei01,tes01,bur02,geb02,mcl03}.  Here, we simply compare
the component spectra of 2MASS 1707-0558 to previously observed SpeX
data for the optical late M and L dwarf spectral standards
\citep{bur04,cruz06}.  Figure 3 overlays the normalized 2MASS
1707-0558 component spectra with those of the comparison stars to
demonstrate the best spectral type fit for each component.  The
northern component spectrum matches the M9 optical spectral standard
LHS 2924 \citep{pro83,kir01} quite well, with equivalent depths of the
TiO and VO bands, and maintaining a similar shape over the entire
spectrum.  The southern component spectrum exhibits diminished TiO and
VO features and increased H$_2$O absorption, matching that of the L3
field dwarf 2MASS J08472872-1532372 \citep{cru03}.  While M and L
spectral type classifications are generally defined at optical
wavelengths, the excellent agreement between the 2MASS 1707-0558 near
infrared spectra and those of the spectral comparators suggest that
our classifications are consistent with the optical types and accurate
to within 0.5 subclass.

An alternative method for near infrared classification is the
calculation of spectral indices of diagnostic regions of the spectrum
which are correlated with spectral types.  We measured the
\citet{rei01} H$_2$O$^{A}$ and H$_2$O$^{B}$ indices, sampling the 1.3 and
1.5$\micron$ steam bands, respectively; and the K1 index
\citep{tok99}, a probe of H$_2$O absorption at 2.0$\micron$.  We then
used the spectral type calibrations given in \citet{rei01},
appropriate over optical spectral types M8 to L6, to derive
classifications of M9 and L2 for the brighter and fainter components,
respectively, with an uncertainty of 2 spectral types, as determined
by the dispersions in each of the linear spectral type fits.
Therefore, the spectral indices confirm the spectral types assigned by
the spectral comparison method, but the large errors of this
classification method prohibits a more precise estimate of the
spectral types.

\subsection{Are 2MASS 1707-0558AB Gravitationally Bound?}

The angular proximity of the 2MASS 1707-0558 components as well as
their similar brightnesses and spectral types, infer that the two
components are gravitationally bound.  One of the more conclusive
tests of physical association is the detection of common space motion.
We examined the astrometry of the system over 3 epochs spanning 2.47
years to determine if the pair shared common proper motion.  Table 2
lists the astrometric results from the PSF fitting, as well as the
expected astrometric results if the secondary object were an
unrelated, unmoving background source.  Figure 4 displays the expected
separation and position angle of the system over time if the system is
not gravitationally bound, with overlying data points for the
measurements from the PSF fitting.  A consistent separation and
position angle over time confirms common proper motion.  The proper
motion of 2MASS 1707-0558, $\rho$ = 0$\farcs$100$\pm$0$\farcs$008 at
88$\pm$10$\degr$ (from the SuperCosmos Sky Survey;
\citet{ham01a,ham01b,ham01c}), is small and implies angular motion of
only 0$\farcs$25 over our observational baseline, about 2 pixels at
the SpeX plate scale.  Our observations are not precise enough to
establish common motion through the angular separation of the two
components; however, our position angle measurements (accurate to
2-3$\degr$) can rule out a stationary background source at the
2.9$\sigma$ (99.6\% confidence) level.

We compare our spectral types and photometric relative magnitudes
derived by PSF fitting to the predicted relative magnitudes for an M9
and L3 binary system.  \citet{cru03} produces a polynomial
M$_J$/spectral type relation (based on 2MASS photometry) using a
sample of ultracool dwarfs (M6-L8) identified in the 2MASS catalog
with independent parallax measurements.  Assuming the stars are
located at the same distance and of spectral type M9 and L3, the
empirical polynomial fit gives $\Delta$M$_{J}$=1.20$\pm$0.35,
consistent with our value of $\Delta$M$_{J}$=1.71$\pm$0.15.  There is
no parallax measurement for 2MASS 1707-0558, but spectrophotometric
distances can be estimated by comparing the apparent magnitudes of
each component to the absolute magnitudes typical for M9 and L3
dwarfs, as calculated using low mass absolute magnitude/spectral type
relations \citep{dah02,cru03,vrb04}.  The two observed sources in this
system lie at a mean distance of 15 pc, with a 1 pc standard deviation
in our measurements.
 
But what if these two sources are nearby but unaffiliated late-type
dwarfs?  We also find that the chance alignment of two similarly
classified low mass stars within 1$\arcsec$ of each other is highly
improbable.  \citet{cru03} performed a wide field search for VLM
stars, discovering 30 M9-L3 stars over an area of 16,350 deg$^{2}$ at
a depth of 20 pc, indicating a space density for these stars of $\rho
\approx$ 0.002 pc$^{-3}$.  At a distance of 15$\pm$1 pc and an angular
separation of 1$\arcsec$, we estimate the volume of space occupied by
the components of 2MASS 1707-0558 to be $V$ $\approx$ 3.3$\times$10$^{-8}$ pc$^3$.
Assuming Poisson statistics, the probability ($P$) that one or more
random low mass stars would be located within this volume of space can
be calculated as
\begin{equation}
P = 1-e^{-\rho V},
\end{equation}
or 10$^{-10}$, ruling out chance alignment with high confidence.  All
the tests discussed in this section lead us to conclude that the
components of 2MASS 1707-0558 comprise a gravitationally bound system.

\section{Discussion}

\subsection{2MASS 1707-0558AB System Characteristics}

We calculate the projected physical separation of 2MASS 1707-0558AB to
be 15$\pm$3 AU, based on the astrometric measurements and derived
spectrophotometric distance.  Temperatures for the two components can
be derived using the T$_{eff}$/spectral type relation for low mass
stars from \citet{gol04}, yielding T$_{eff}$ = 2400$\pm$175 K for
2MASS 1707-0558A and 1950$\pm$190 K for 2MASS 1707-0558B.  We
calculate individual masses using the \citet{burr97} models for ages
of 0.5 and 5 Gyr, sampling the typical ages for field dwarfs (see
below).  The derived masses are 0.072-0.083 M$_{\sun}$ and 0.064-0.077
M$_{\sun}$ for the A and B components, respectively, with a total
system mass of 0.136-0.160 M$_{\sun}$ and a mass ratio $q \equiv$
M$_2$/M$_1$ = 0.88-0.92.  This mass ratio is again consistent with the
properties of VLM binaries, which are predominantly near-equal mass
systems.  Combining the estimates of the physical separation and mass
indicates an orbital period of roughly 150-300 yr, an unfortunately
unreasonable timescale for dynamical mass measurements.  A summary of
the 2MASS 1707-0558AB system characteristics can be found in Table 4.

\subsection{Broader Implications for M and L Dwarfs}

This binary system is similar to other field binary systems, with a
small separation and nearly equal mass components.  93\% of known VLM
binaries have separations less than 20~AU, like 2MASS 1707-0558, and
77\% have $q \geq 0.8$ \citep{bur06b}.  Therefore, this system further
strengthens the case that VLM binaries have specific traits that may
be related to their formation mechanism.

The M9/L3 binary 2MASS 1707-0558 is also a useful probe of the M/L
transition, since the components have presumably coevolved with a
fixed age and metallicity.  Resolved binaries with spectroscopy are
ideal systems to study the complex atmospheric chemistry of low mass
stars, especially the thickening of dust, effects of condensation on
the spectral energy distribution, and the relative heights of the
cloud base.  M dwarfs are characterized by their TiO and VO bands
\citep{kir99}, while the transition into the L dwarf regime is marked
by the depletion of these molecules through condensation, leading to
the expression of hydrides such as FeH and CrH \citep{lod02}.  The
atmospheric chemistry has been modeled for an assortment of field
stars, yielding various masses, ages, and intrinsic metallicities.  A
study of resolved binary systems straddling the M/L boundary, such as
2MASS 1707-0558, will permit a fair comparison of low mass
atmospheres.

In addition, comparisons of the individual rotational velocities
through atmospheric line broadening may constrain particular formation
and evolution models.  A direct measurement would determine if the
components have equivalent or varied rotation rates, clarifying a
dependency on either the mass or the origin of the object.
Chromospheric activity is not well understood in low mass stars, but
it is likely to be correlated to the spin rate of the star, possibly
as a result of the magnetic fields induced by a turbulent dynamo
mechanism \citep[see also \citealt{cha06}]{dur93}.  Field
studies suggest that H$\alpha$ activity increases from early-type to
late-type M dwarfs, and then diminishes through the early-type L
dwarfs \citep{giz00,moh03,wes04}.  An unresolved optical spectrum of
2MASS 1707-0558 shows modest H$\alpha$ emission \citep{giz02}, but the
origins of this detection are unclear.  Is this activity present in
both components, and at what level does each component contribute to
this detection?  Furthermore, a measurement of the soft X-ray
contributions of each component (resolvable with {\em Chandra}) will
help further constrain the magnetic field mechanisms and their effects
on coronal heating.

The substellar nature of brown dwarfs can be directly tested through
measurements of the 6708 {\AA} \ion{Li}{1} line, but unfortunately the
original optical spectrum was unable to resolve this line.  The
existence of primordial lithium is a conclusive determinant of
substellar nature, as it is quickly destroyed when core temperatures
reaches $\sim$ 2$\times$10$^{6}$ K, a prerequisite reached by VLM
stars with masses greater than $\sim$0.06M$_{\sun}$
\citep{reb92,mag93}.  It should be noted that an object can lie below
the hydrogen-burning minimum mass (i.e., substellar) and fail to
exhibit \ion{Li}{1} absorption.  Depending on the age of the system,
either one or both of the components will show \ion{Li}{1} in
absorption since they follow different evolutionary tracks.  Hence, in
the situation of a binary such as 2MASS 1707-0558 that may straddle
the substellar boundary, the existence of the \ion{Li}{1} line can be
also used to trace the age of the system (cf., \citealt{liu05}).
Three possible scenarios exist: 2MASS 1707-0558A and B could both be
brown dwarfs and contain \ion{Li}{1} undepleted from their atmospheres
($t$ $\le$ 0.5 Gyr), 2MASS 1707-0558A could be a star or brown dwarf
with mass greater than $\sim$0.06M$_{\sun}$ and 2MASS 1707-0558B a
Li-bearing brown dwarf with a mass less than $\sim$0.06M$_{\sun}$ (0.5
Gyr $<$ $t$ $<$ 1 Gyr), or both components could be low mass stars or
brown dwarfs with masses greater than $\sim$0.06M$_{\sun}$ ($t$ $>$ 1
Gyr).  2MASS 1707-0558 is therefore a rare VLM field system that can
be assigned an accurate age through the \ion{Li}{1} diagnostic.

An alternative age diagnostic, although less reliable, is a
consideration of space kinematics and chromospheric activity.  Stars
are generally born with the space motion of their natal cloud, and
over time accumulate higher space velocities through interactions with
the components of the Galactic disk \citep{wei77}.  \citet{giz00} have
found that a sample of field M and L dwarfs separate into two groups
according to kinematics and chromospheric activity.  An old population
of M dwarfs have high tangential velocities (v$_{tan}$ $>$ 20 km
s$^{-1}$) and strong H$\alpha$ emission, while a younger population
exhibits low kinematics and low activity.  Assuming our
spectrophotometric distance and the proper motion as measured by the
SuperCosmos Sky Survey, 2MASS 1707-0558 has a V$_{tan}$ of 7 km
s$^{-1}$.  In comparison to the other late M dwarfs identified in
\citet{giz00}, 2MASS 1707-0558 has one of the lowest tangential
velocities and H$\alpha$ measurements, suggesting that 2MASS 1707-0558
is associated with the young population.  It is therefore likely to be
closer to $\sim$1~Gyr in age, but without additional constraints, we
adopt a conservative age range of 0.5-5~Gyr.  Figure 5 compares the
properties of the 2MASS 1707-0558 components to evolutionary models
from \citet{burr97}, clearly showing the secondary's possible location
below the Li-burning mass limit.

\section{Summary}

We have photometrically and spectroscopically resolved the 2MASS
1707-0558 into a M9/L3 binary separated by 1$\farcs$01$\pm$0.17.
Physical association is confirmed by common proper motion, angular
proximity, similar distances, and the statistical likelihood of two
low mass stars occupying this small volume of space.  System
characteristics were derived by combining the imaging and spectral
information obtained with the IRTF SpeX instrument, and considering
the current evolutionary models and empirical large scale surveys.
The derived $J$,$H$,$K$-band relative magnitudes are consistent with
assigned spectral classifications and imply a spectrophotometric
distance of 15$\pm$1 pc.  The properties of 2MASS 1707-0558AB,
including close separation and high mass ratio, are typical for late
M/L binary systems.  The angular separation of this system enables
resolved spectroscopic measurements critical for the studies of
low-mass star formation, atmospheric chemistry, and activity across
the M/L transition.

\acknowledgments

We thank our telescope operators Bill Golisch, Dave Griep, and Paul
Sears, and instrument specialist John Rayner, for their support during
the IRTF observations.  We would also like to thank James Larkin, Stan
Metchev, and Peter Plavchan for many useful conversations regarding
the science presented herein, Michael Cushing for discussions on
synthetic photometry, and our anonymous referee for a careful review
of the manuscript.

This publication makes use of data from the Two Micron All Sky Survey,
which is a joint project of the University of Massachusetts and the
Infrared Processing and Analysis Center, funded by the National
Aeronautics and Space Administration and the National Science
Foundation.  2MASS data were obtained from the NASA/IPAC Infrared
Science Archive, which is operated by the Jet Propulsion Laboratory,
California Institute of Technology, under contract with the National
Aeronautics and Space Administration. The authors wish to extend
special thanks to those of Hawaiian ancestry on whose sacred mountain
we are privileged to be guests. Electronic copies of the spectra
presented here can be obtained directly from the primary author.

\clearpage

\begin{figure}
\epsscale{1.0}
\plotone{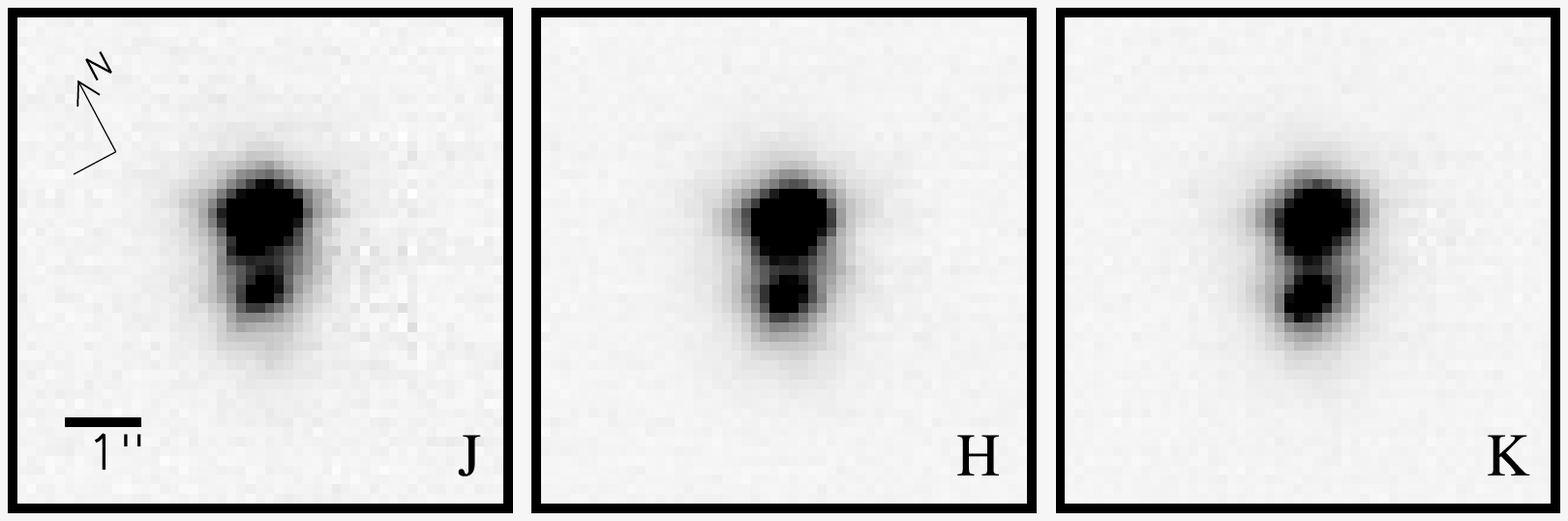}
\caption{The reduced mosaic discovery images of 2MASS 1707-0557AB in
the $J$- (left), $H$- (middle), and $K$-bands (right) obtained on 2003
March 23.  Images are 6$\arcsec$ on a side, with North (arrow) and
East (line) indicated in the first panel.}
\label{fig1}
\end{figure}


\begin{figure}
\epsscale{1.0}
\plotone{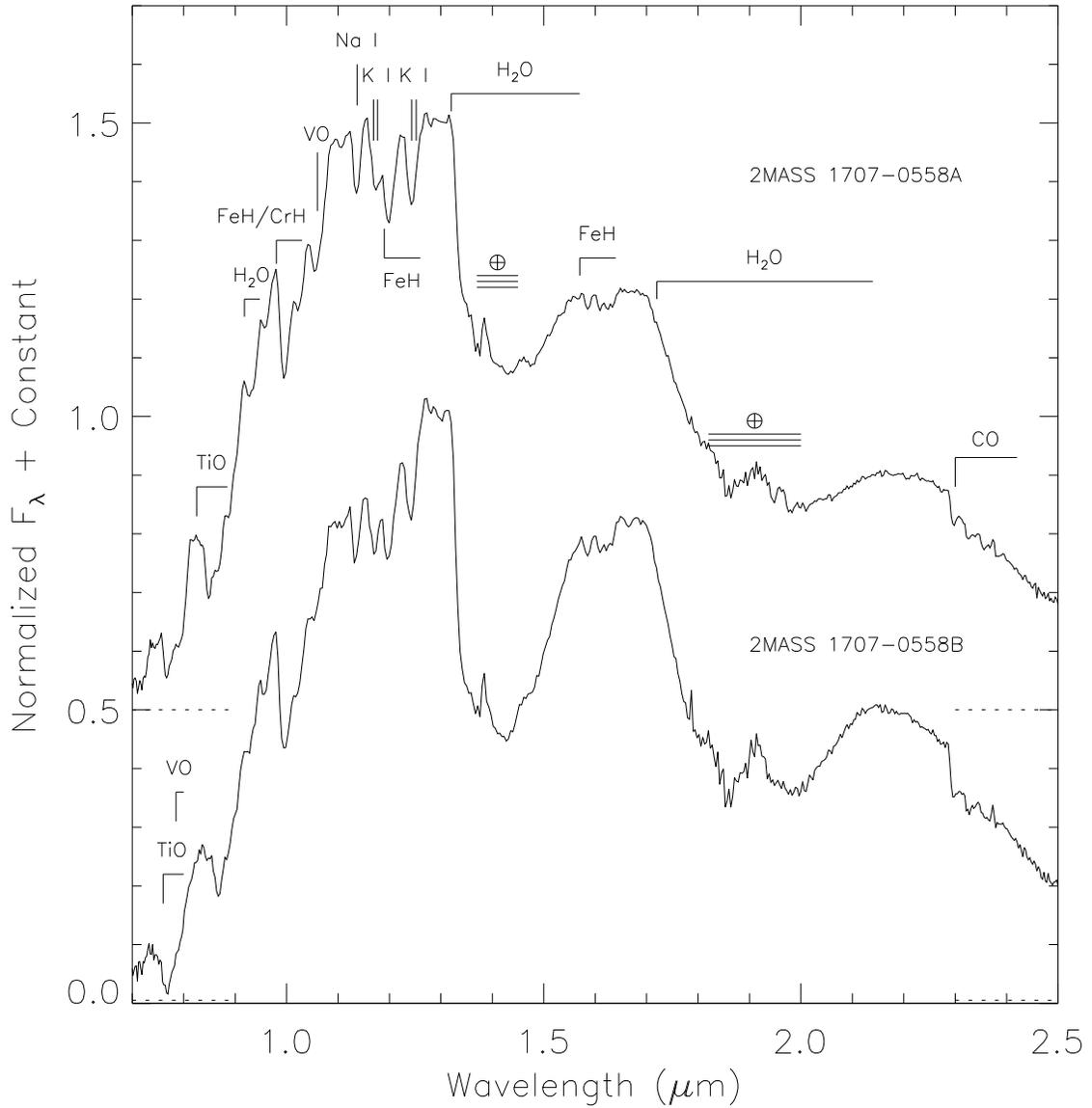}
\caption{Near infrared SpeX spectra of 2MASS 1707-0558A (top) and B
(bottom), with the major absorption features of TiO, VO, H$_{2}$O,
FeH, CrH, \ion{Na}{1}, and \ion{K}{1} indicated, as well as regions affected by
telluric absorption (circled-plus symbols).  All data are normalized
at 1.27$\micron$ and offset by a constant.  The zeropoint of each
spectrum is designated by a dotted line.
\label{fig2}}
\end{figure}


\begin{figure}
\epsscale{1.0}
\plotone{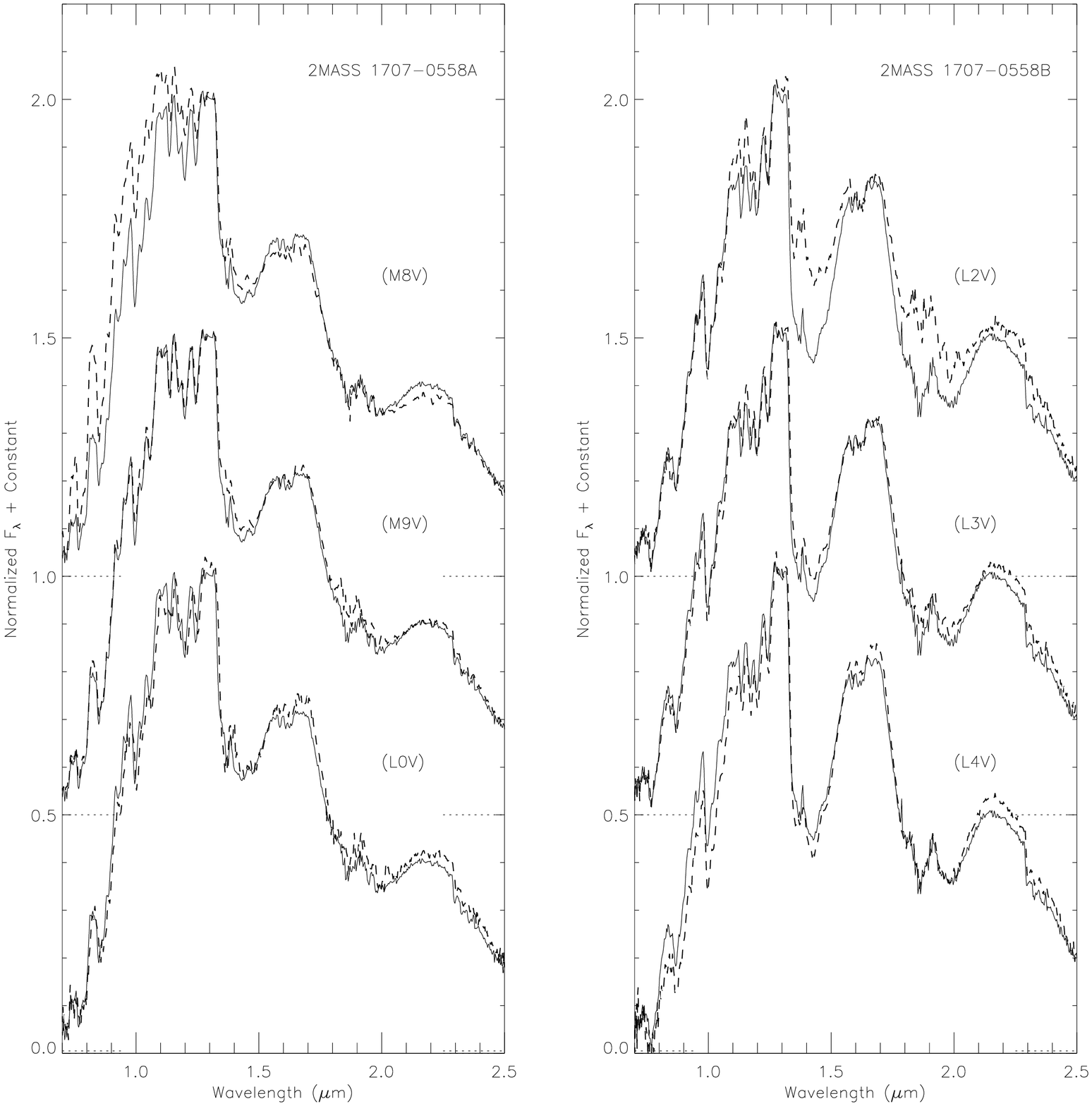}
\caption{Comparison of 2MASS 1707-0558A and B spectra (solid lines) to
optically defined spectral standards (dashed lines).  ({\em Left})
2MASS 1707-0558A overlaid on the spectral standards VB 10
\citep[M8;][]{pro83,kir01}, LHS 2924 \citep[M9;][]{pro83,kir01}, and
2MASS 0345+2550 \citep[L0;][]{kir97}.  ({\em Right}) 2MASS 1707-0558
overlaid on 2MASS 0847-1532 \citep[L2;][]{cru03}, SDSS 2028+0052
\citep[L3;][]{haw02}, and 2MASS 1104+1959 \citep[L4;][]{cru03}.  All
data are normalized at 1.27$\micron$ and offset by a constant.}
\label{fig3}
\end{figure}


\begin{figure}
\epsscale{1.1}
\plottwo{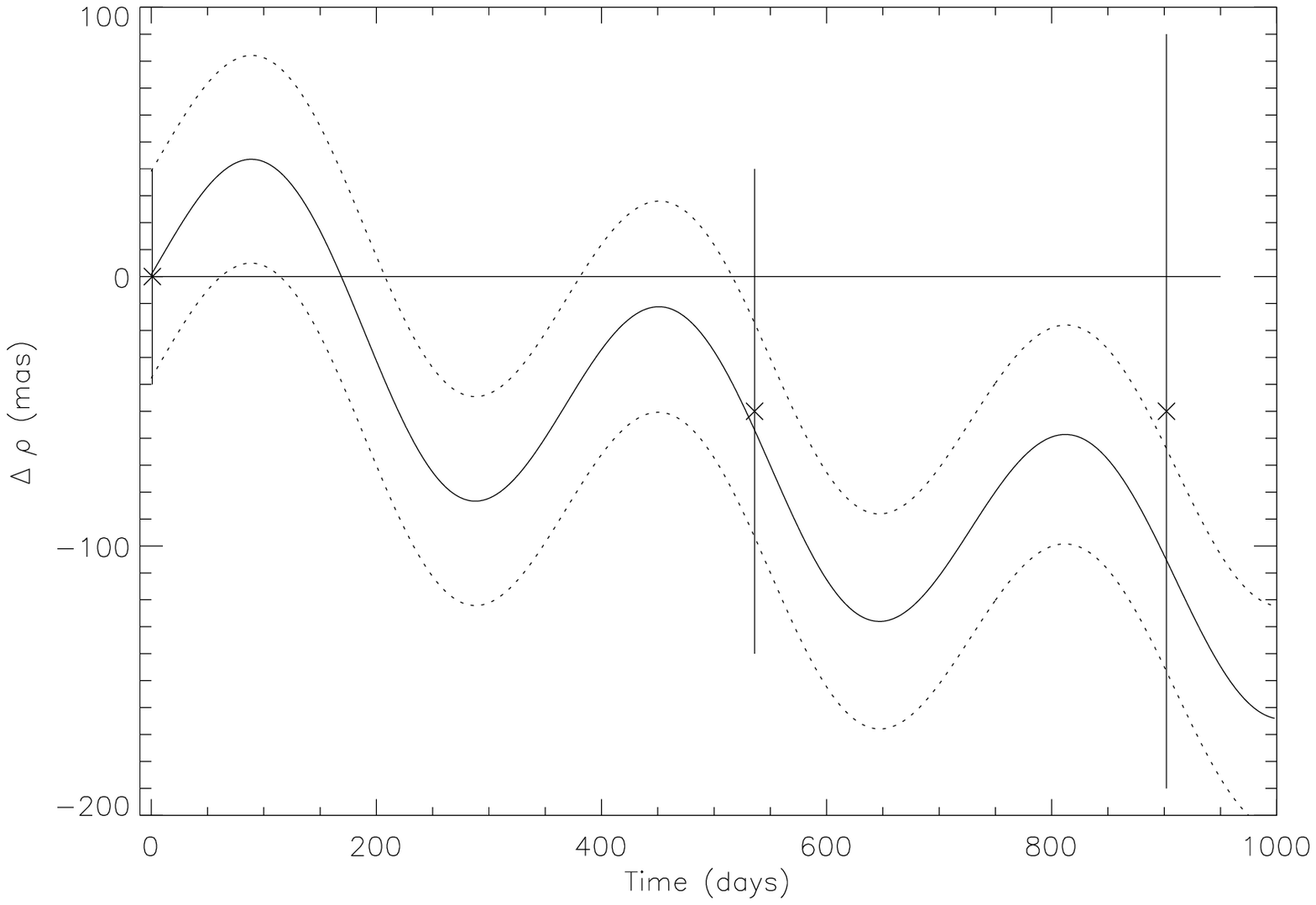}{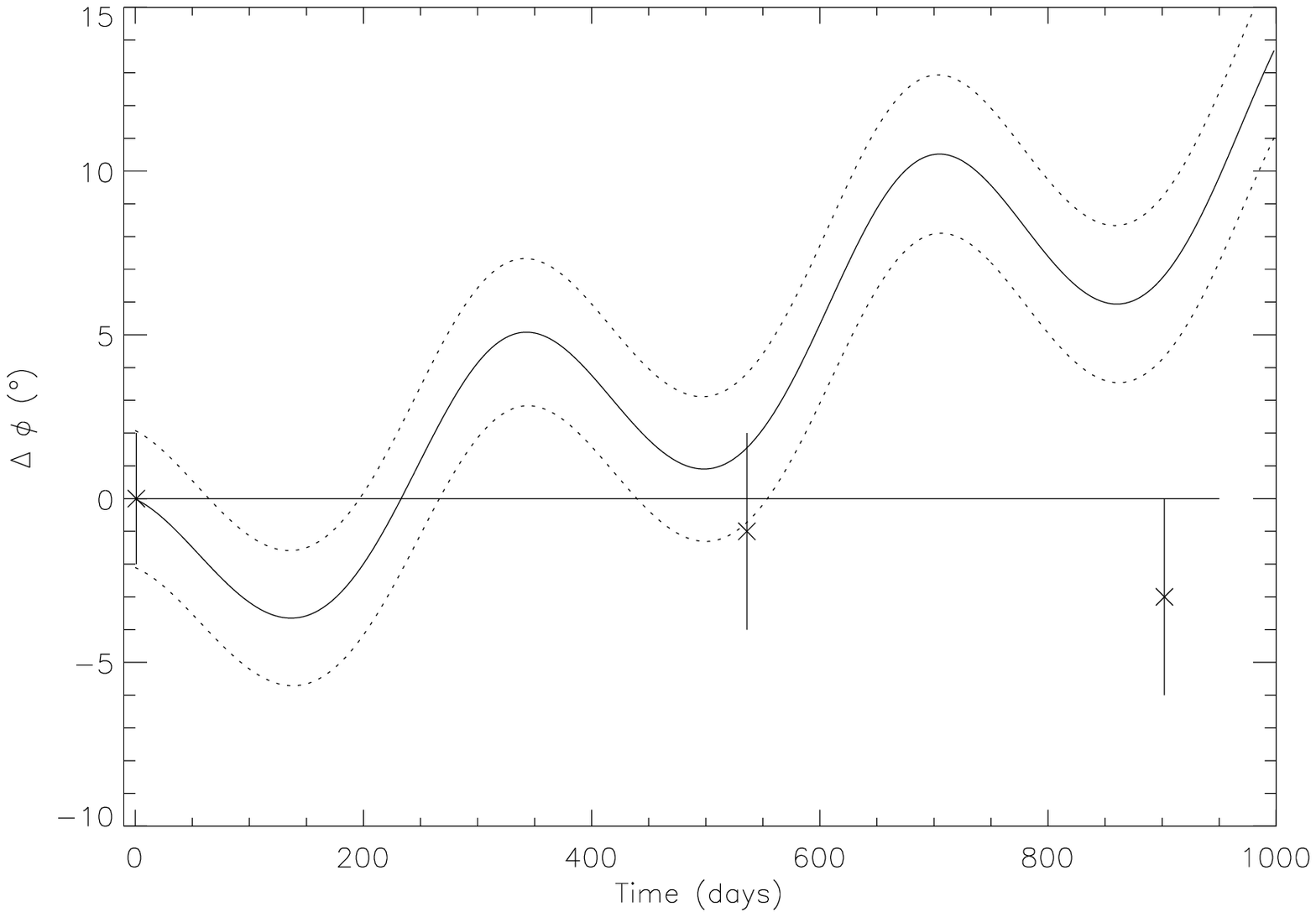}
\caption{Astrometric motion of 2MASS 1707-0558B relative to A.  Curves
show the angular separation ({\em left}) and position angle ({\em
right}) expected for an unmoving, unassociated background source, with
dotted lines corresponding to the 1$\sigma$ errors in the estimated
positions.  The straight line marks the position of a companion
sharing common proper motion with the primary.  Data points are the
mean values from the PSF fitting algorithm, with the errors
corresponding to the 1$\sigma$ uncertainty in these measurements.  The
periodicity of the relative background source position is due to the
parallactic motion of the primary at its estimated spectrophotometric
distance (15$\pm$1 pc).}
\label{astrometry}
\end{figure}


\begin{figure}
\epsscale{1.0}
\plotone{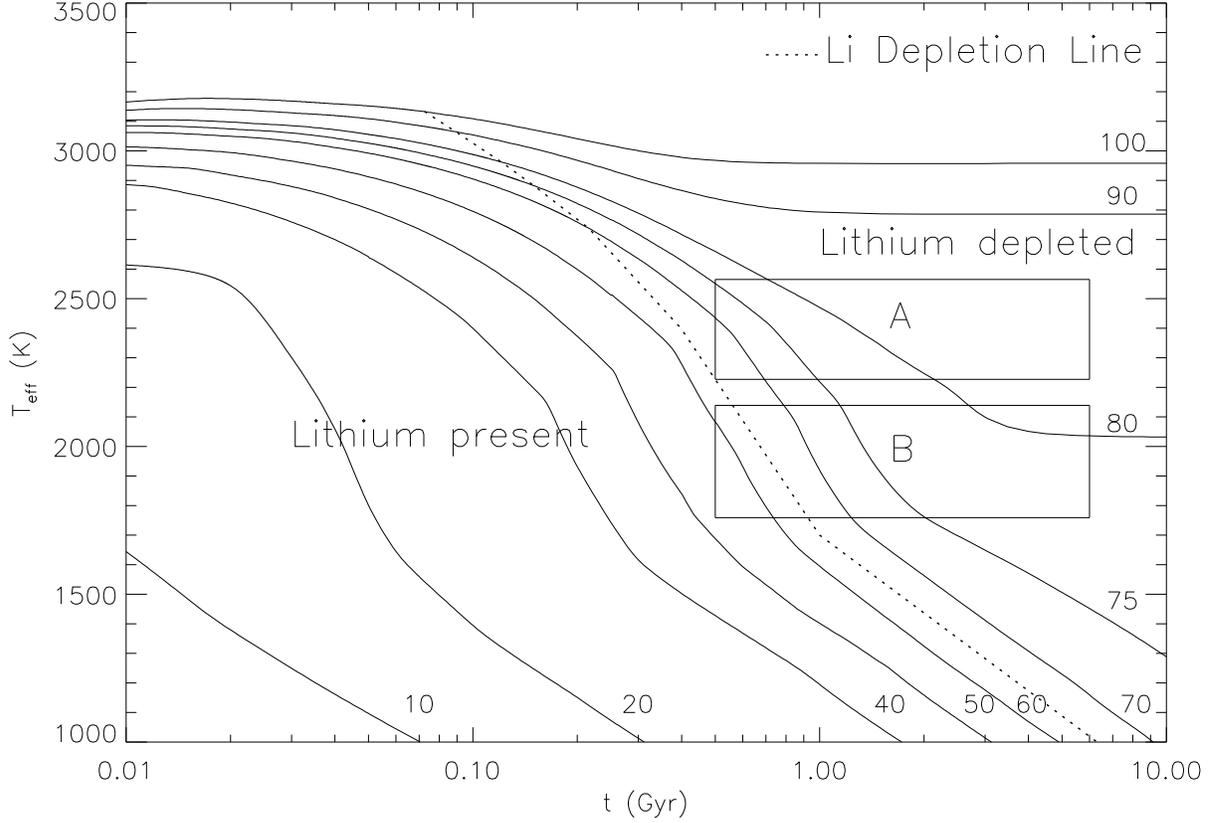}
\caption{T$_{eff}$ evolution as computed by Burrows et al. (1997,
solid lines) for masses of 10, 20, 40, 50, 60, 70, 75, 80, and 100
M$_{Jup}$ (1 M$_{Jup}$ $\approx$ 0.001 M$_{\sun}$).  The tracks are
labeled from left to right and bottom to top, respectively.  The
mass/age loci of the 2MASS 1707-0558 A and B components are indicated,
based on T$_{eff}$s derived using the empirical T$_{eff}$/spectral
type relation of \citet{gol04} and an age estimate of 0.5-5~Gyr.  The
L3 (B) component is likely to be a brown dwarf, while the M9 (A)
component is located near the hydrogen-burning minimum mass.  The
dotted line displays the boundary where more than 90\% of the
primordial lithium abundance is depleted, according to the models of
\citet{burr97}.}
\label{bur_evol}
\end{figure}

\clearpage

\begin{deluxetable}{lccccc}
\tablenum{1}
\label{indices}
\tabletypesize{\scriptsize}
\tablecaption{Observing log.}
\tablewidth{0pt}
\tablehead{
\colhead{UT Date} & \colhead{Observation} & \colhead{t$_{int}$ (s)} &
\colhead{Airmass} & \colhead{$J$-band Seeing ($\arcsec$)} & \colhead{Calibrator} \\
}
\startdata
2003 March 23 & $JHK$ imaging &  60, 60, 60 & 1.17 & 0.5 & none \\
& 0.8-2.5$\micron$ spectroscopy & 480 (A), 720 (B) & 1.16 & 0.5 & HD 171149 \\
2004 August 9 & $JHK$ imaging & 80, 60, 48 & 1.17 & 0.9 & USNO-A2.0 0825-10078125 \\
2004 August 10 & $JK$ imaging & 40, 40 & 1.15 & 0.9 & USNO-A2.0 0825-10079794 \\
\enddata
\end{deluxetable}

\clearpage

\begin{deluxetable}{lcccc}
\tablenum{2}
\label{astrometry}
\tabletypesize{\scriptsize} \tablecaption{2MASS 1707-0558 System
Astrometry.}  \tablewidth{0pt} \tablehead{
\colhead{} & \multicolumn{2}{c}{Measured Astrometry} & \multicolumn{2}{c}{Expected Astrometry} \\
\colhead{} & \multicolumn{2}{c}{} & \multicolumn{2}{c}{If Background Source} \\
\colhead{Date} & \colhead{Offset} & \colhead{PA} & \colhead{Offset} & \colhead{PA}\\ 
\colhead{} & \colhead{(arcsec)} & \colhead{(deg)} & \colhead{(arcsec)} & \colhead{(deg)}\\ } 
\startdata
2003 March 23 & 1.04$\pm$0.04 & 145$\pm$2 & $\cdots$  & $\cdots$\\
2004 August 9 & 0.99$\pm$0.09 & 144$\pm$3 & 0.97$\pm$0.04 & 147$\pm$2\\
2004 August 10 & 0.99$\pm$0.14 & 142$\pm$3 & 0.94$\pm$0.04 & 152$\pm$2 \\
\enddata
\end{deluxetable}

\clearpage

\begin{deluxetable}{lcc}
\tablenum{3}
\label{sys_props}
\tabletypesize{\scriptsize} \tablecaption{2MASS 1707-0558AB System
Properties.}  \tablewidth{0pt} \tablehead{ \colhead{Parameter} &
\colhead{Value} & \colhead{Ref} \\ } 
\startdata
$\alpha$\tablenotemark{a} & 17$^{h}$07$^{m}$23$^{s}$.43 & 1 \\
$\delta$\tablenotemark{a} & $-$05$^{\degr}$58$\arcmin$24$\farcs$9 & 1 \\ 
$\mu$ & 0$\farcs$100$\pm$0$\farcs$008 yr$^{-1}$ & 2 \\ 
$\theta$ & 88$\pm$10$\degr$ & 2 \\ 
d\tablenotemark{b} & 15$\pm$1 pc & 3 \\ 
$\rho$ & 1$\farcs$01$\pm$0$\farcs$17 & 3 \\ & 15$\pm$3 AU\tablenotemark{c} & 3 \\ 
$\phi$ & 145$\pm$3$\degr$ & 3 \\ 
J\tablenotemark{d} & 12.05$\pm$0.02 mag & 1 \\
H\tablenotemark{d} & 11.26$\pm$0.03 mag & 1 \\ 
K\tablenotemark{d} & 10.71$\pm$0.02 mag & 1 \\ 
$\Delta$J & 1.71$\pm$0.15 mag & 3 \\
$\Delta$H & 1.14$\pm$0.10 mag & 3 \\ 
$\Delta$K & 1.18$\pm$0.12 mag & 3 \\ 
M$_{tot}$\tablenotemark{e} & 0.136-0.160 M$_{\sun}$ & 3,4 \\
q\tablenotemark{e} & 0.88-0.92 & 3,4 \\ 
Period\tablenotemark{e} &
$\sim$150-300 yr & 3,4 \\ 
\enddata 
\tablenotetext{a}{Equinox J2000 coordinates at epoch 1999 April 09 from 2MASS.}
\tablenotetext{b}{Spectrophometric distances derived using absolute
magnitude/spectral type relations; see $\S$~3.3.}
\tablenotetext{c}{Physical separation derived from the angular separation and the spectrophometric distance.}
\tablenotetext{d}{2MASS photometry of the unresolved system.}
\tablenotetext{e}{Assuming an age of 0.5-5 Gyr} \tablerefs{(1) 2MASS
\citep{skr06}; (2) SuperCosmosSky Survey \citep{ham01a,ham01b,ham01c};
(3) This paper; (4) \citet{burr97}.}
\end{deluxetable}

\clearpage

\begin{deluxetable}{lcc}
\tablenum{4}
\label{indprops}
\tabletypesize{\scriptsize}
\tablecaption{2MASS 1707-0558AB Individual Properties.}
\tablewidth{0pt}
\tablehead{
\colhead{} & \colhead{2MASS 1707-0558A}  & \colhead{2MASS 1707-0558B} \\
\colhead{Parameter} & \colhead{Value} & \colhead{Value} \\
}
\startdata
SpT & M9\tablenotemark{a} & L3\tablenotemark{a}\\
T$_{eff}$\tablenotemark{b} (K) & 2400$\pm$175 & 1950$\pm$190\\
J\tablenotemark{c} & 12.26$\pm$0.11 & 13.96$\pm$0.11\\
H\tablenotemark{c} & 11.59$\pm$0.07 & 12.72$\pm$0.04\\
K$_{s}$\tablenotemark{c} & 11.03$\pm$0.08 & 12.20$\pm$0.08\\
Estimated Mass\tablenotemark{d} (M$_{\sun}$) & 0.072-0.085 & 0.064-0.077\\
\enddata
\tablenotetext{a}{Spectral type uncertainty of $\pm$0.5 subclasses.}
\tablenotetext{b}{Derived from the T$_{eff}$/spectral type relation of \citet{gol04}.}
\tablenotetext{c}{Magnitude on the 2MASS photometric system.}
\tablenotetext{d}{Estimated masses from 0.5 to 5 Gyr, based on the evolutionary models of \citet{burr97}.}
\end{deluxetable}

\clearpage

\end{document}